\documentclass[runningheads]{llncs}
\usepackage{graphicx}
\usepackage{hyperref}
\usepackage{booktabs}
\usepackage{multirow}
\usepackage{paralist}
\usepackage{xcolor,colortbl}
\usepackage{cite}
\usepackage{amsmath}
\usepackage{nicefrac}
\usepackage{todonotes}
\usepackage{xspace}

\definecolor{Gray}{gray}{0.9}
\newcolumntype{g}{>{\columncolor{Gray}}r}
\newcolumntype{h}{>{\columncolor{Gray}}c}

\begin{document}

\title{Bridging HPC and Quantum Systems using Scientific Workflows
\\
}

\author{
    Samuel T.\ Bieberich\orcidID{0000-0001-9631-647X}\inst{1,2} \and
    Ketan C.\ Maheshwari\orcidID{0000-0001-6852-5653}\inst{2} \and
    Sean R.\ Wilkinson\orcidID{0000-0002-1443-7479}\inst{2} \and
    Prasanna Date\orcidID{0000-0002-1664-069X}\inst{2} \and
    In-Saeng Suh\orcidID{0000-0002-6923-6455}\inst{2} \and
    Rafael Ferreira da Silva\orcidID{0000-0002-1720-0928}\inst{2}
}
\authorrunning{S.\ T.\ Bieberich et. al.}

\institute{
    Texas A\&M University, College Station, TX, USA \and
    Oak Ridge National Laboratory, Oak Ridge, TN, USA\footnote{\scriptsize This manuscript has been authored by UT-Battelle, LLC, under contract DE-AC05-00OR22725 with the US Department of Energy (DOE). The publisher acknowledges the US government license to provide public access under the DOE Public Access Plan (http://energy.gov/downloads/doe-public-access-plan).}
    \\
    \email{
        sambieberichtamu@tamu.edu, \{maheshwarikc,wilkinsonsr,datepa,suhi,silvarf\}@ornl.gov
    }
}

\maketitle

\begin{abstract}
Quantum computing offers intriguing potential to solve certain kinds of
problems with unprecedented speed. Quantum computers are unlikely to replace
classical computers in the future, but 
may work in tandem with them to perform complex
tasks by utilizing their complementary strengths. Indeed, most quantum
computers today are made available to users via cloud-based Application
Programming Interfaces (APIs) which must be called remotely from classical computers.
Unfortunately, this usage model presents obstacles for a seamless application execution connecting quantum
computers with classical High Performance Computing (HPC) systems. Workflow management systems can help overcome these obstacles.

In this work, we apply the scientific workflows paradigm to bridge the gap
between quantum and classical computing -- specifically, between the quantum
and HPC systems available through the Oak Ridge Leadership Computing Facility
(OLCF). We provide three fully automated foundational examples for demonstration: the
Traveling Salesman Problem, Grover's Search Algorithm, and Shor's Factoring
Algorithm. We employ workflows to generate inputs from OLCF's HPC systems and
transfer them to IBM Quantum systems in the cloud, where the quantum calculations
produce results which return to OLCF for post processing.

This workflows-based approach provides additional benefits including
\emph{(a)} end-to-end programmatic automation of the entire process, 
\emph{(b)} an out-of-the-box tool for interfacing with HPC schedulers and
           quantum middleware, and
\emph{(c)} concurrency of independent tasks such as running the same algorithm
           over a simulator and a real quantum device simultaneously.
Although the current technological limitations of quantum computers prevent the
use of these algorithms to solve real-life problems at scale, the
workflows-based approach nevertheless unites these two powerful computing
paradigms in a way that shows immense promise for the future.

\end{abstract}


\keywords{High Performance Computing, Quantum Computing, Scientific Workflows}
\section{Introduction}
Five decades after its initial statement, Moore's Law~\cite{moores-law} is
arguably ``dead'' because the rate at which transistor technology can continue
to be miniaturized is now limited by the physical limits of transistors
themselves~\cite{end-of-moores-law}. The density of transistors on a computer
chip has long been a strong indicator of the performance of the chip. Taken
together, these facts would seem to foreshadow grave consequences for the
advancement of classical computing -- and by extension, High
Performance Computing (HPC). Advances in quantum computing have been
encouraging for the HPC community, which is ever in need of greater
computational performance.

While some of the contemporary quantum computing offerings provide an API to
interact with the various layers of a quantum ecosystem, none provide an
interface that could simultaneously interact with both traditional HPC and
quantum computing. Workflow management systems are very well adapted to
traditional HPC systems~\cite{ferreiradasilva2021works}. For instance, most workflow systems natively
provide features to automatically perform HPC job management portably while
others provide interfaces for reliable data transfer. As such, workflow systems
are well-suited to provide and develop ``attachments'' to a new system. In this
work, we demonstrate how a workflow system may be used to bridge the gap
between HPC and quantum computing platforms.

Despite being a promising frontier of computing, quantum computers, in their current 
state-of-the-art, have several limitations. For example, current superconducting 
quantum processors need to be kept extremely cold around absolute zero temperatures, 
and even the largest universal quantum computers have less than 433 qubits 
available~\cite{ibmroadmap}. Moreover, these chips are unable to allow for all of the 
qubits to be entangled directly, thereby severely limiting connections between certain 
qubits. This results in two specific limitations. First, today's quantum computers 
solve problems that are very small in size -- this is practical for experimental 
purposes only. Second, any internal and external noises can cause ``errors" into the 
quantum system causing programs to often give incorrect outputs and requiring them 
to be run multiple times in order to obtain results that are inferenced 
statistically.

\begin{figure}[!t]
    \centering
    \vspace{-5pt}
\includegraphics[width=\linewidth]{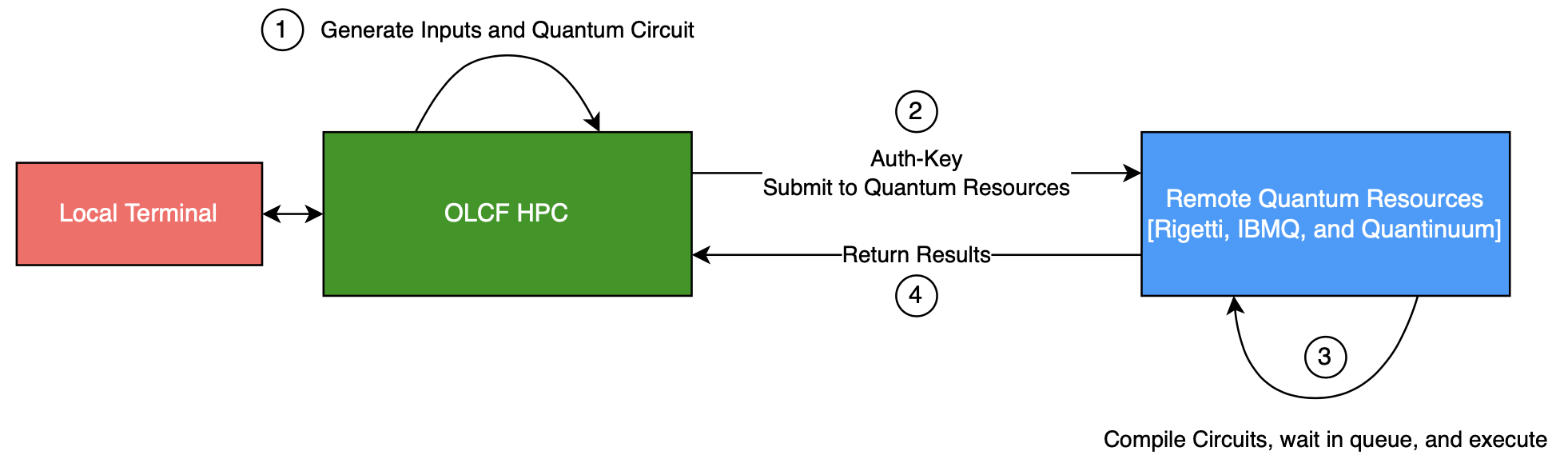}
    \vspace{-20pt}
\caption{Overall workflow schema between tradional HPC and Quantum Systems.}
\label{fig:workflow}
\end{figure}

Figure~\ref{fig:workflow} shows the overall schema of our setup. A local
terminal gets connected to a login node of the HPC system (which could potentially 
be replaced by a graphical user interface such as Jupyter notebooks). 
A workflow system that is running on the HPC system generates the inputs and a
preprogrammed quantum circuit. The workflow system invokes the quantum
resources by submitting the circuit and input data with appropriate
authentication credentials. The quantum system performs the necessary
compilations and computing and produces the outputs. The workflow system that
is interfaced with the appropriate API collects the results and brings them
back at the HPC site.  The code and other implementation artefacts are publicly
available on Github~\cite{Bieberich_HPC-QC-Workflows_2022}. 

On the HPC side, we use Crusher, a precursor to the upcoming Frontier
supercomputer, and Andes, a commodity cluster at OLCF for the experiments
presented in this paper. Crusher~\cite{crusher} is OLCF's moderate-security
system that contains identical hardware and similar software as the Frontier
system (the first exascale HPC system). It is  used as an early-access testbed
for the Center for Accelerated Application Readiness (CAAR) and Exascale
Computing Project (ECP) teams as well as OLCF staff and the vendor partners.
The system has 2 cabinets, the first with 128 compute nodes and the second with
64 compute nodes, for a total of 192 compute nodes. Each compute node is
equipped with 64-core AMD EPYC 7A53 ``Optimized 3rd Gen EPYC" CPU, four AMD
MI250X, each with 2 Graphics Compute Dies (GCDs) for a total of 8 GCDs per node
with access to 64 GiB of HBM2E, 512 GB of DDR4 memory, and connection to a 250
PB GPFS scratch filesystem.

The rest of the paper is organized as follows. Section~\ref{sec:tsp} describes
the algorithm, implementation, and workflow-bridging schema for the Traveling
Salesman Problem. Sections~\ref{sec:grover} and~\ref{sec:shor} describe the
Grover's search and Shor's factoring algorithms, respectively. Both the 
implementations follow the same workflow-bridging schema as described in 
section~\ref{sec:tsp}. Section~\ref{sec:related} describes the related work 
from both research and industry. Finally, section~\ref{sec:concl} presents 
our conclusions and future work.

\section{Traveling Salesman Problem}
\label{sec:tsp}

\begin{figure}[!t]
    \centering
    \vspace{-5pt}
\includegraphics[scale=0.2]{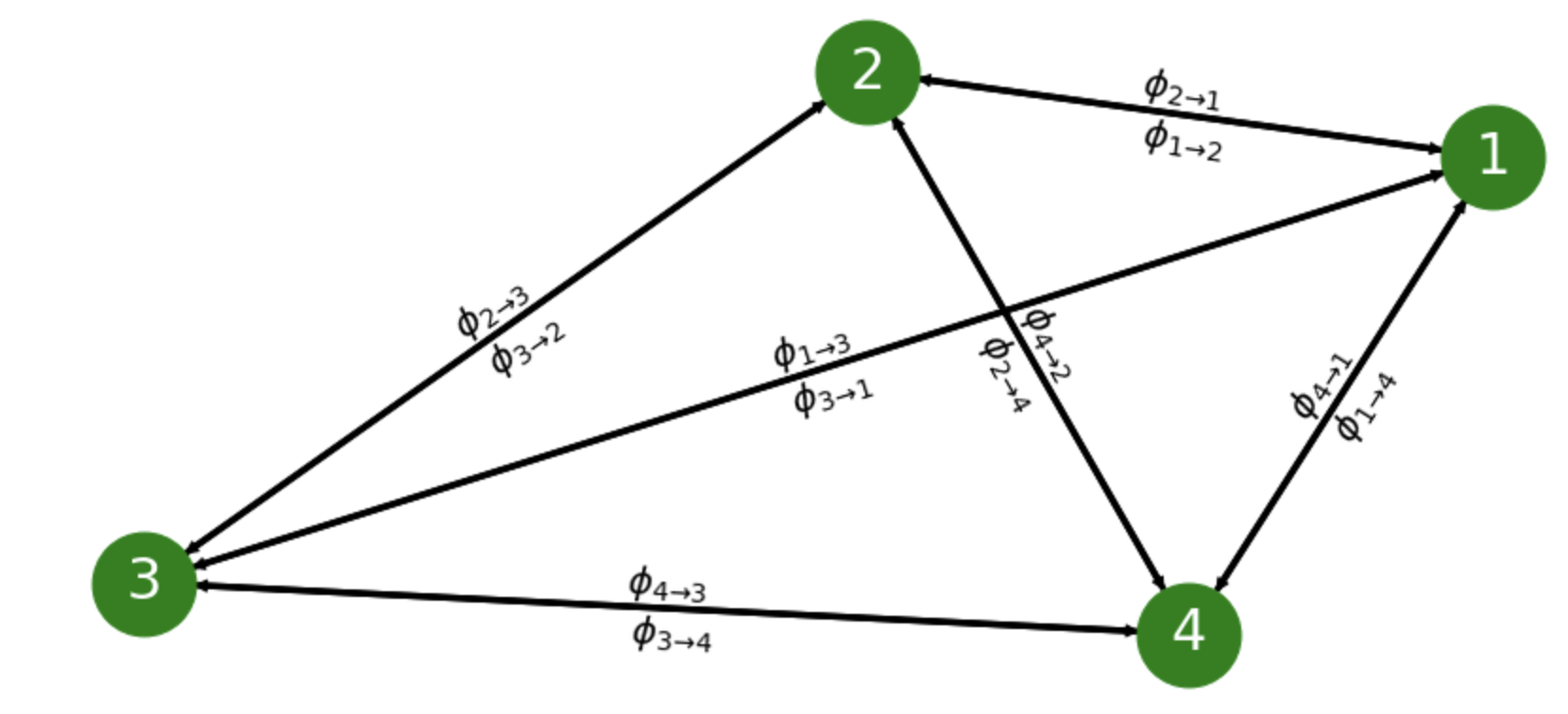}
    \vspace{-10pt}
\caption{Example NetworkX map for a four-node TSP.}
\label{fig:tsp-map}
\end{figure}

The Traveling Salesman Problem (TSP) is a well known fundamental optimization 
problem with significant practical importance. Classified as an NP-Hard 
problem, the TSP is not solvable in polynomial time, meaning as more nodes 
are added, the problem gets exponentially harder for classical 
computers~\cite{junger1995traveling}. The TSP asks for the fastest way to 
visit a number of cities $N$ (also referred to as nodes), given the 
distances between them, and make it back, while traveling the shortest 
distance. This results in $(N-1)!$ different routes that may be taken. While 
there are many current algorithms to solve TSP implementations with 
relatively few nodes, even the largest supercomputers are unable to find the 
best distance with hundreds of nodes in polynomial time. 
Figure~\ref{fig:tsp-map} shows a an example of a randomized map for four 
cities generated using NetworkX~\cite{hagberg2008exploring}.

The TSP is not hard to compute with a calculator at four nodes, much less a 
supercomputer, however the limited access to powerful quantum computers led 
us to create the circuit for only four nodes. Each time our code was 
implemented on HPC, one of these NetworkX maps was imported to our local 
laptops, and when the quantum jobs were completed, we checked the answers. 
Roughly rectangular maps such as the one pictured in figure~\ref{fig:tsp-map} 
often lead to two paths with very similar distances. This error is accepted 
in current TSP algorithms on classical computers (albeit with much more 
nodes), and thus we assumed that TSPs with that particular shape could have 
two sufficient correct paths.

\subsection{Algorithm}

The process for designing the TSP 4-node circuit utilizes phase estimation.
Phase estimation is a method that allows for users to read information about an
operation from qubits in
superposition~\cite{johnston_harrigan_gimeno_segovia_2019}.  We initially
attempted to use Quantum Approximate Optimization Algorithms (QAOA), but these
algorithms did not offer the exact answer we were pursuing with the limited
access to quantum
qubits~\cite{bergamaschi_2020,radzihovsky_murphy_swofford_2019}.  It is worth
noting that these algorithms could be better for very large TSP
problems~\cite{bergamaschi_2020}. 

The algorithm we used utilized matrices that would be brute forced by 
classical computers and converted to phases. These phases are often 
represented on the Bloch Sphere, a well-known 3D representation of how 
qubits physically work, and are synonymous with rotations around an axis 
of the sphere. After getting unitary matrices for each of the four nodes, 
these can be converted to high-level gates in a quantum algorithm, 
primarily composed of Controlled-Not (CNOT), Rotation, and SWAP gates. 
A concurrent step involves determining the Eigenstates. As aforementioned, 
there are $(N-1)!$ paths in a TSP, where N is the number of nodes being 
tested. Each path can be mapped to a unique Eigenstate, which for the rest 
of the program must be represented in binary, so both the classical and 
quantum computers can read it. The paths are converted to binary Eigenstates 
via the function $i(j)$, which defines the TSP. 

\begin{equation}
    |\psi\rangle = \otimes_j |i(j)-1\rangle 
\end{equation}

For example, in the path 1-2-3-4-1, if the path from node 2 to node 3 is 
taken, then $i(3) = 2$, thus $j$ is the number for the node you are 
traveling to. After taking the tensor product for 
each value 1 through 4, the Eigenstate is completed. To further optimize 
the program, it can be proven that the paths 1-2-3-4-1 and 1-4-3-2-1 are 
the same, thus the number of Eigenstates can be reduced to $(N-1)!/2$, or 
3 for a four-node TSP.

\subsection{Implementation}

\begin{figure}[!t]
	\centering
	\vspace{-5pt}
{\includegraphics[width=\linewidth]{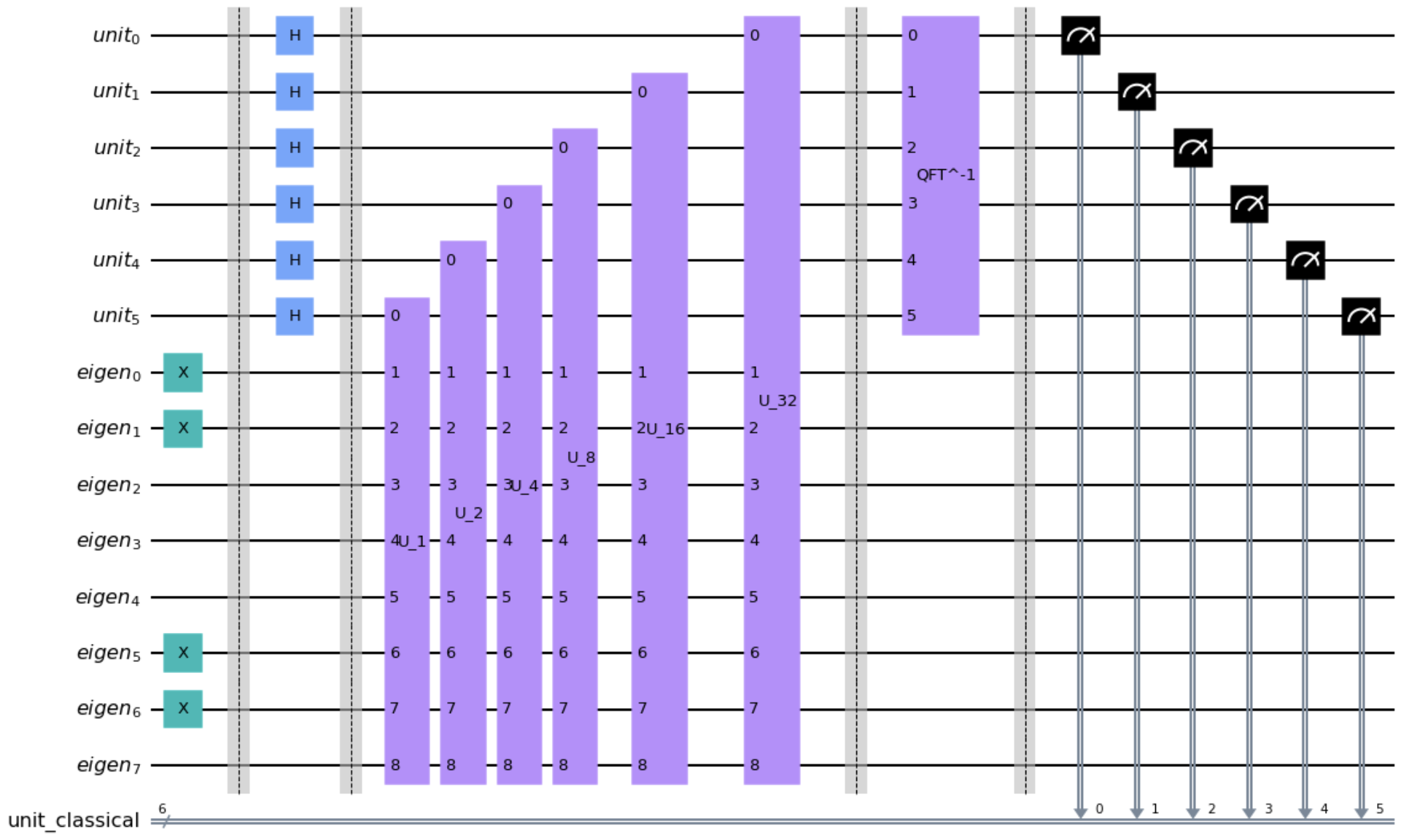}}
    \vspace{-20pt}
\caption{Circuit for 4 node TSP with Eigenstate 11000110, or path 1-2-3-4-1}
\label{fig:tsp-circuit}
\end{figure}

Figure~\ref{fig:tsp-circuit} shows the high-level gates for the 4 node TSP 
circuit. Using the Phase Estimation method, there are two registers, the 
Unit and Eigenstate (shortened to eigen in figure~\ref{fig:tsp-circuit}) 
registers. The quantum part of the algorithm can be split into four parts. 
The Unit register is initialized with Hadamard Gates, putting each qubit 
into a superpositioned state. The Eigen register, as expected, is initialized 
based on the Eigenstate the circuit is testing. Because there are three 
Eigenstates, there are three circuits in this algorithm that are tested 
altogether. Also, because the Eigenstates in binary are eight digits long, 
the register is composed of 8 qubits. Going from index 0 to 7 of the binary 
Eigenstate, an X gate is applied to the initialization step of the Eigen 
register at the corresponding Qubit index. X gates are fundamentally identical 
to Boolean NOT gates, and flip the intial states of the qubits from 0 to 1, 
offering an ``input" to the TSP circuit. 

The second step of the process is the actual Phase Estimation. As explained 
above, the Phase Estimation part of the program can be decomposed into matrices 
which convert the Eigenstates to, in conjunction with the QFT, eigenvalues, 
which can be read by a computer. Phase Estimation circuits are composed of 
Controlled-Unitary (CU) gates, which accept a control value from the unit 
register, and if the control value is measured as a 1, the Unitary in the 
Eigen register runs~\cite{johnston_harrigan_gimeno_segovia_2019}. There are 
the same number of CU gates as there are qubits in the unit register. It is 
worth mentioning that the unit register, unlike the Eigen register, can be 
increased or decreased in size. The code from Qiskit's Alpha Textbook used 
6 qubits, and through testing, we determined that adding more qubits only 
marginally increased accuracy, and decreasing the qubits sacrificed accuracy
(see Section~\ref{sec:results}). 

The third step of the process is the Inverse Quantum Fourier Transform. The 
$QFT^{-1}$ finishes the conversion from Eigenstates to eigenvalues, and 
prepares the unit register to be measured. Lastly, the results are measured 
to a classical register.

\subsection{HPC-Quantum Workflow}

In this section, we describe the overall workflow for TSP. The workflow scheme 
remains the same for the other alforithms, however, it was arguably most 
complex for TSP. After completing the code and basic sanity tests over local 
hardware, we uploaded the Jupyter Notebook code into HPC enabling rapid 
prototyping. For each algorithm, we needed to save our IBMQ accounts with our 
unique API tokens. After this code ran, we were able to use the ``load account" 
function from the Qiskit library to load our credentials, rather than leaving 
the long API token in the file. We did this due to security concerns, as we did 
not want to have to change our API token very often, as then each Jupyter 
notebook would need to be adjusted for our testing with the other algorithms 
running concurrently. 

After loading the credentials, the rest of the code was split into five main 
steps:
\begin{itemize}
\item The code to create random coordinates for each of the four nodes and then 
      graphing them into the TSP format. A file is created that sends a picture 
      of the node map to the local computer. 
\item A python function reads the coordinates from the map and finds the 
      distances between each node. Another, overarching function converts these 
      distances into a matrix, then a python list so that it can interface 
      correctly with the Controlled-U gate creation function.
\item The rest of the circuit is built via Qiskit, including the inverse QFT, 
      which to conserve gates and allow for scalability, was automatically 
      produced from the Qiskit library in the TSP program. 
\item The three circuits are sent together, as a list, to IBMQ. Each time, 
      \emph{ibmq kolkata} was the least busy processor, so each test (besides 
      the simulations) was run on it. We ran the circuits four times, each 
      taking three to six hours in the queue and approximately 2 and a half 
      minutes to run, with 4,000 shots (default value).
\item The outputs are read via two variables: The most frequent counts are 
      determined, and functions find which path is the shortest and verify that 
      it is right. The results are printed in the terminal. IBMQ creates a 
      histogram on their web portal, allowing for a more readable graph than 
      the Qiskit histograms that are readable from the terminal. 
\end{itemize}

The entire aforementioned workflow was automated using Parsl~\cite{parsl}, a 
popular workflow management tool. In addition to automating the workflow, 
Parsl allows for python functions to run concurrently to increase the speed 
of implementation. We were able to complete the workflow with several steps 
running concurrently. For instance, the NetworkX map process takes several 
seconds, so we organized it to run concurrently with the quantum circuit 
initialization. We were able to run the circuit on the IBMQ QASM simulator 
and ibmq\_kolkata quantum computer \emph{simultaneously}, allowing for 
predictive results from the simulator to generate as the circuit awaits in the 
queue for the real quantum computer. We refer readers to our Github codebase 
for further implementation details~\cite{Bieberich_HPC-QC-Workflows_2022}.

\subsection{Results}
\label{sec:results}

\begin{figure*}[!t]
    \centering
    \vspace{-5pt}
    \includegraphics[width=\linewidth]{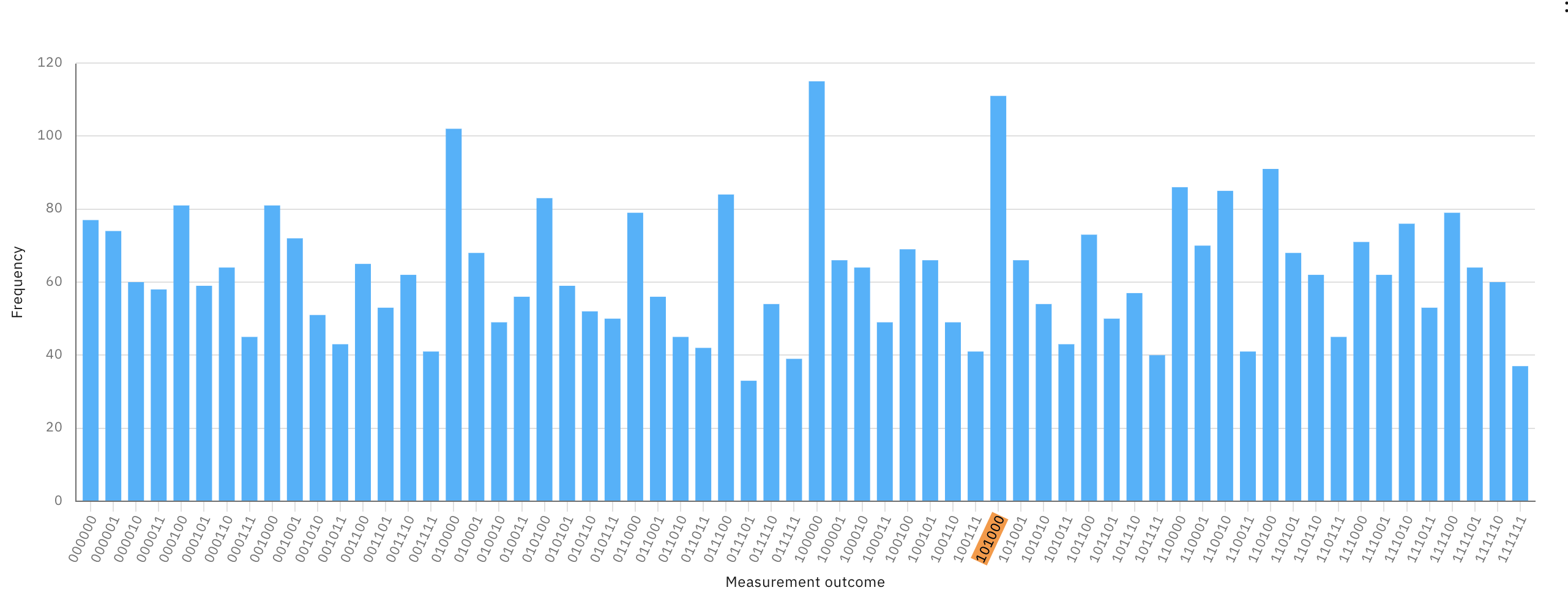}
    \vspace{-20pt}
    \caption{1-2-3-4 path TSP circuit results from ibmq\_kolkata. Correct answer highlighted.}
    \label{fig:tsp-ibm}
\end{figure*}

The results are printed as binary strings that are 6 digits long (one for each 
qubit). Due to the nature of the Phase Estimation, the largest numbers 
represent the shortest paths that may be taken. For example, in one test we 
ran, the largest value was 35, which was returned from circuit one, associated 
with Eigenstate 11000110, or 1-2-3-4. This means that the shortest path through 
the TSP is 1-2-3-4-1. 


In terms of measured results, we tested the circuit first through HPC and 
IBMQ's ibmq\_kolkata 27 qubit cloud quantum computer, then via local and 
IBMQ's ibmq\_qasm\_simulator, which can operate up to 32 qubits. The real 
quantum computer, as exhibited in Figure~\ref{fig:tsp-ibm}, features a 
significant amount of noise, rendering the results inconclusive. The 
correct result for the TSP generated with this run should be 101000, which 
has a greater frequency than many of the other measured values, however it 
is not the highest, and almost every measured value is represented far too 
much. Due to the size of the circuit, too much noise was likely introduced, 
resulting in significant error. 

On the other hand, the ibmq\_qasm\_simulator proved to offer much more 
accurate results. With the same path, 1-2-3-4-1, the true result of 101000 
was consistently returned. The reason this result was so much more accurate 
has to do with the composition of the QASM simulator. The QASM simulator runs 
on a classical computer via the Cloud, and while it does model noise, it 
supports all of the gates in the circuit we wrote, meaning the compiler step 
does not need to split the CU gates into thousands of simpler gates, rather it 
only has hundreds. It ignores the calibration issues that modern quantum 
computers need, and assumes that all gates are connected, decreasing gate count 
by several magnitudes.  Full results, along with results for the Grover's and 
Shor's algorithms code, are publicly available 
online~\cite{Bieberich_HPC-QC-Workflows_2022}.

\section{Grover's Search Algorithm}
\label{sec:grover}

Grover's algorithm finds an item in an unsorted list of length $N$ using only 
$O(\sqrt{N})$ operations, as opposed to $O(N)$ operations on average for a 
classical computer~\cite{adedoyin2018quantum}. In a nutshell, the algorithm 
corresponds to a bar graph with one bar representing each index of the 
computational list~\cite{Qiskit-Textbook}. The oracle in the Grover's algorithm 
finds the value being searched for and flips it from a value of 1 to -1. Then, 
it finds the average of all of the values in the list, and flips each index 
over said value. This way, the index at which the value is at has a 
significantly larger magnitude than all other indices, thus making it easy to 
identify.

\subsection{Implementation}

Quantum programs in IBMQ's Qiskit programming language are composed of quantum 
circuits. These circuits are composed of a variety of quantum gates, similar in 
concept to Boolean Logic gates in Digital Computers. These circuits are read 
from left to right, and are composed like musical staves, with one qubit 
represented by each horizontal line. 

\begin{figure}[!t]
    \centering
    \vspace{-5pt}
    \includegraphics[width=\linewidth]{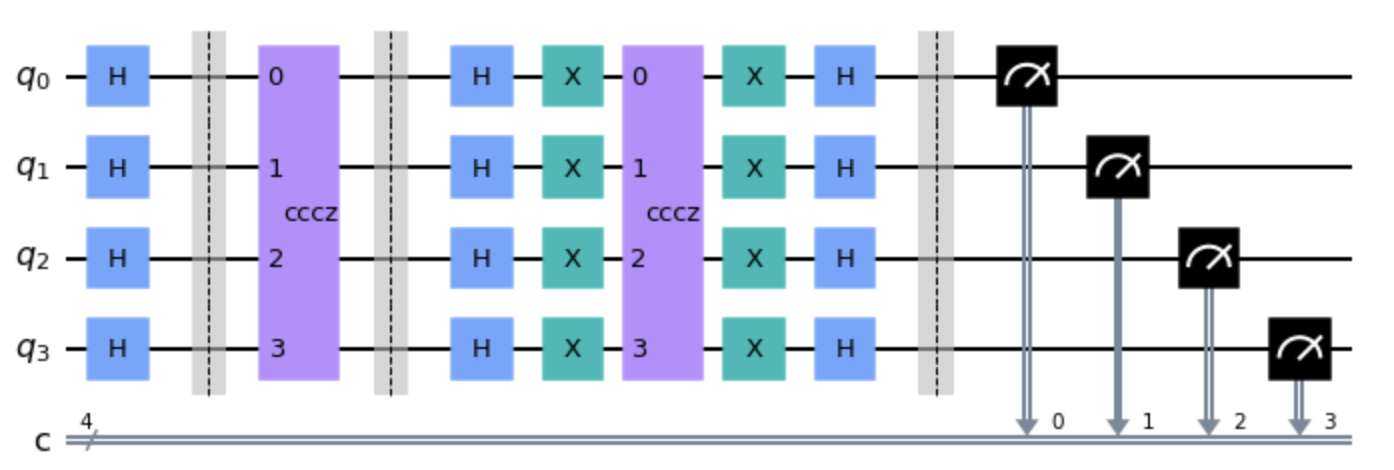}
    \vspace{-20pt}
    \caption{IBMQ Matplotlib circuit for Grover's Algorithm, with one implementation of the Grover Operator. Value being searched for is 15.}
\label{fig:grover}
\end{figure}

Figure~\ref{fig:grover} shows an example of the most basic part of Grover's 
Algorithm for the integer value 15. (Each section is separated by barriers for 
formatting purposes.) The first section uses Hadamard (H) gates to put each 
qubit into superposition. The second section is a manually created 
Controlled-controlled-controlled-Z gate (CCCZ), with controls on qubits 0-2 
and a Z gate applied to q3. This section of the circuit is the Oracle, and 
changes depending on the value input. Oracles are often referred to as ``Black 
Boxes", and are created in most cases by the processor. The rest of the program 
is designed to help you discover what the oracle is. If the value were 0, each qubit would have X (NOT) gates applied on either side of the CCCZ gate. The third section is the Amplification function. This is the part of the algorithm where each value is flipped over the average of all indices~\cite{Karlsson20184qubitGA}. Lastly, the fourth and final section of the circuit measures each qubit. Each of the four measurement gates convert the qubits from their quantum register to a classical register, which a regular computer can then read as a 1 or 0. Since there are four qubits, the integer values 0-15 can be returned. 

Grover's Algorithm is split into three steps when explained, step one being Initialization, then a Grover Operation, then Measurement. The second and third sections of our circuit are combined into one statement, the Grover Operator. Grover's Algorithm is most accurate when the Grover Operator is repeated $\sqrt{N}$ times, where N is the number of qubits. Since our implementation of Grover's Algorithm has 4 qubits, the Grover Operator must be completed twice before measurements are made. 

\subsection{Results}
We designed the quantum circuit and the rest of the program in Jupyter Notebooks with the IBMQ's Qiskit~\cite{Qiskit}. To run the program via a workflow encompassing HPC and quantum cloud computers, we first uploaded the completed program on the HPC side by adding it to a GitHub repository and then pulling the code to a terminal. The program randomly chose a number between 0 and 15. Once this number was chosen, it was printed and saved as a variable (to compare with results), then a loop created the unique Grover Oracle. Once this was complete, the circuit was sent to IBM's ibmq-belem 5 qubit quantum computer. After 1024 shots of the circuit, the results are brought back to the HPC side, which makes a histogram displaying the results. On average, 92\% of the shots returned the number input. 

\section{Shor's Factoring Algorithm}
\label{sec:shor}
More than any other quantum algorithm, Peter Shor's factoring algorithm has created the most buzz for physicists and computer scientists. Current encryption techniques, such as the prevalent RSA encryption in everything from governmental to financial resources, are composed of keys made of the factors of RSA-2048, a 2048 bit number. These factors are still unknown, and almost impossible to find, because they are both prime numbers. This makes RSA-2048 a semi-prime integer, the hardest to factor, as it is divisible by nothing but those two numbers. 

Shor's algorithm is composed of a series of steps, starting with a classical
computer, then transferring a circuit to a quantum computer, and finally
reading the results on a classical computer to determine if the circuit needs
to be run on the quantum device again with a different guessed value. The
process is based on an algorithm that has long been theorized, but is very
difficult to implement at a large scale on regular computers: the Period
Finding Problem. For a given number $N$ that we wish to factor and a randomly selected number $a$ ($1 < a < N$), the period finding problem states that there exists a number $r$
such that $a^r\mod{N} = 1$.
This leads to the greatest common divisor of
$a^{r/2}\pm1$ and $N$ being one of the prime factors of $N$. The steps incurred in the Shor's algorithm are:
\begin{enumerate}
\item Pick a random number $a$ between 1 and $N$, where $N$ is the number being factored.
\item Compute the greatest common divisor (GCD) of $a$ and $N$.
\item If the GCD of $a$ and $N$ is not equal to 1, then $a$ is one of the factors as required. The other factor can be computed by dividing $N$ by $a$.
\item Else, run the quantum period finding subroutine on a quantum computer with $N$ and $a$ as the inputs.
\item Determine the period $r$ by interpreting the results from the quantum period finding subroutine on a classial computer.
\item If the $r$ is 1, redo steps 1--5 with a different value for $a$.
\item If $r$ is odd, restart the process with a different value for $a$.
\item If $r$ is neither $1$ nor odd, compute the GCD of $a^{r/2}\pm1$ and $N$.
\item The GCD should be one of the factors of $N$ as required. Divide $N$ by the GCD returns the second factor as well.
\end{enumerate}

Using these steps, RSA-2048 and other large semi-primes could be factored in the future, though thousands of coherent qubits and an equally large quantum volume would be needed to implement this program for such a large number. Presently, quantum computers can factor semi-primes up to 21 only \cite{adedoyin2018quantum}. 

\subsection{Implementation}

The circuit for Shor's algorithm that we used utilizes two qubit registers, one that encompasses 0-2 qubits, called the work register, and the second encompassing qubits 3-6, called the control register. Only the work register is measured in the end. 

\begin{figure}[!t]
    \centering
    \vspace{-5pt}
\includegraphics[scale=0.4]{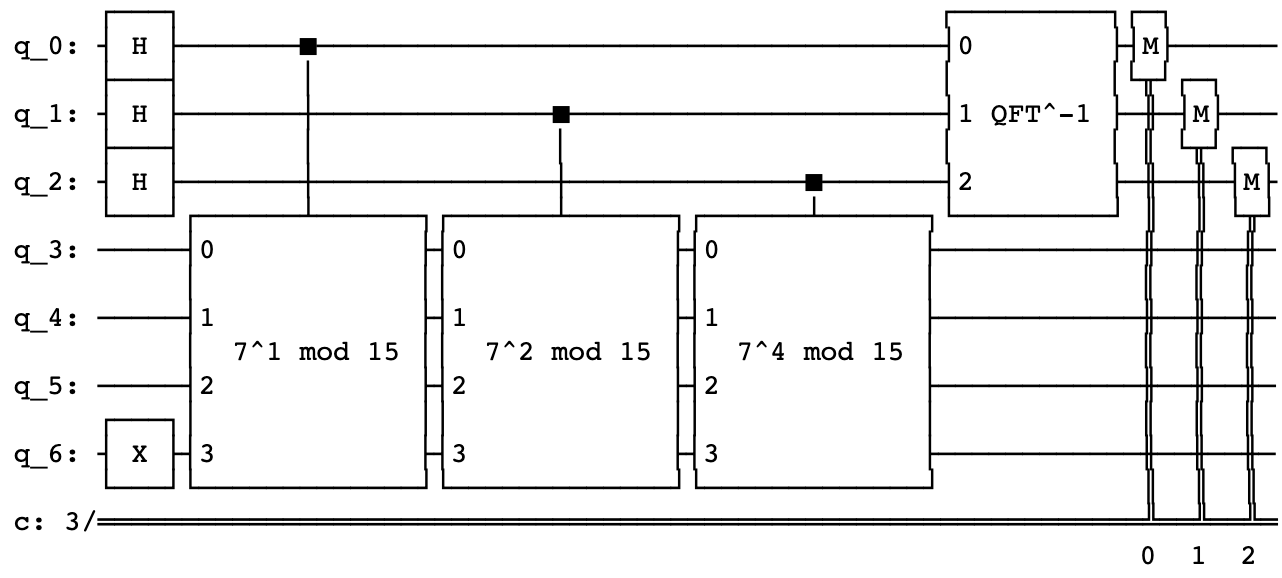}
    \vspace{-10pt}
\caption{IBMQ circuit for Shor's algorithm, with a = 7, on 7 qubits}
\label{fig:shor}
\end{figure}

Like the Grover's algorithm circuit, Shor's circuit can be broken into four distinct sections: Initialization, Modular Exponentiation, Inverse Quantum Fourier Transform ($QFT^{-1}$), and Measurement, as shown in figure~ \ref{fig:shor}. In the Initialization step, all three qubits in the work register are put into superposition, and a NOT gate is applied to the final qubit in the control register. The Modular Exponentiation stage uses $U^{2^j}$ gates to perform a Quantum Phase Estimation on the three work register qubits, resulting in the work register ending in a state a$^x$modN. The Inverse QFT takes these values and creates interference between these states, turning the current circuit value and converting it to a Fourier basis~\cite{amico2019experimental}.

\begin{equation}
    QFT |x\rangle = 1/\sqrt{N} \sum_{y=0}^{N-1} e^{2\pi ixy/N} = |y\rangle
\end{equation}

Lastly, the measurement stage, like in Grover's algorithm, measures the results and sends them to a classical register. Note that the measurement gates only apply to qubits 0, 1, and 2. 

\subsection{Results}

\begin{figure}[!t]
    \centering
    \vspace{-5pt}
\includegraphics[width=\linewidth]{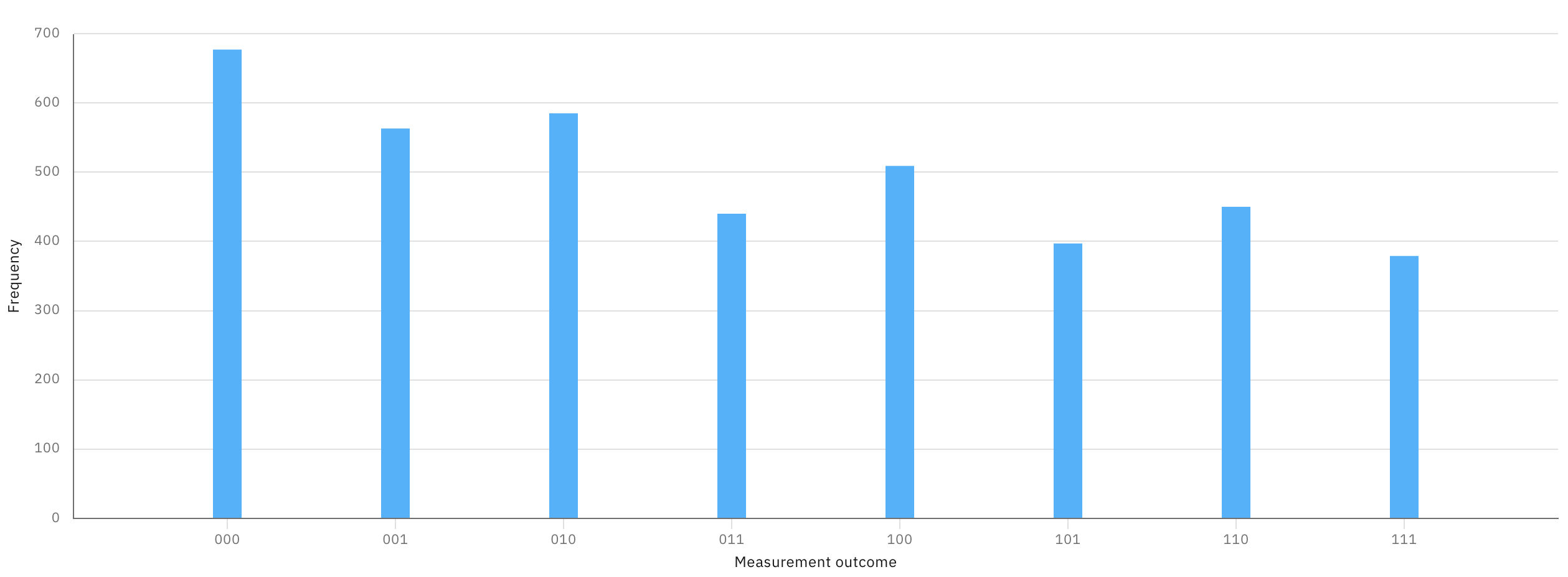}
    \vspace{-20pt}
\caption{Results for Shor's algorithm with a = 13, on ibm\_nairobi}
\label{fig:shor-ibm}
\end{figure}

\begin{figure}[!t]
    \centering
    \vspace{-5pt}
\includegraphics[scale=0.4]{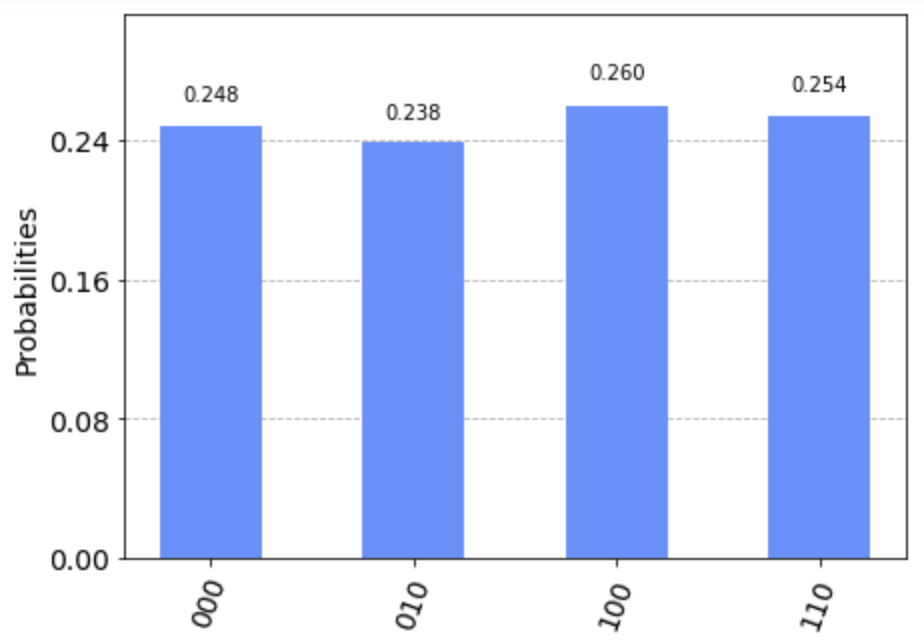}
    \vspace{-10pt}
\caption{Results for Shor's algorithm with a = 13, on IBM's QASM simulator (local) backend}
\label{fig:shor-qasm}
\end{figure}

The larger a circuit in Qiskit is, the more computational time it takes to run and thus more noise is exposed to the delicate qubits running the operations. Thus, the results become further from the expected values, as opposed to the smaller Grover circuit. Figure~\ref{fig:shor-qasm} showing the results from a simulator (IBMQ's qasm\_simulator), and figure~\ref{fig:shor-ibm} showing the same circuit run on a real quantum device, ibm\_nairobi. While figure~\ref{fig:shor-qasm} clearly shows a period of 2, figure~\ref{fig:shor-ibm} is less obvious, and requires knowledge of what the expected answer should be to determine the real period. 

Once again, these circuits were executed from HPC terminal, and results were printed to files we returned to our local computers. Figure~\ref{fig:shor-ibm} is from the IBMQ Jobs recap page on their website, as it offers a less cluttered graph than the default output that can be run via Qiskit. This graph has an almost perfect four-way distribution of the measured counts: 0, 2, 4, and 6, in binary. The way the circuit was designed, 0 always features a large distribution (such as in~\ref{fig:shor-ibm}), and is thus ignored, but the following values, 2, 4, and 6, exhibit the period of the function.

\section{Related Work}
\label{sec:related}
Several works have researched the ways to implement quantum workflows
for classical-quantum hybrid computing. One recent work experiments with workflows using the D-Wave quantum annealer,
reconciling with the inherent difficulty in creating cloud-based hybrid workflows for
`big science' jobs~\cite{10.1007/978-3-030-50433-5_40}. Additionally, in the
previous editions of ICCS, work such as~\cite{10.1007/978-3-030-77980-1_10},
proposes the ``QuantMe" framework, which models, transforms, and then
computes data via quantum workflows at the research level. It takes into
account the inherent noise in NISQ era quantum computers, and how conversions
from classical to quantum algorithms are affected by this significant
shortcoming in modern technology~\cite{weber_2022}. This helps to mitigate or
eliminate the incredibly technical elements that must take place to set up and
read the results from quantum algorithms, which could slow research such as
ours. Expanding on this theme,~\cite{distributed} detailed the running of a similar
``The Total Weighted Tardiness Problem", which is NP-hard, defines a series of tasks with due dates which must be completed on a
machine, with the goal to determine in which order to complete the tasks to
minimize tardiness. Quantum annealing devices like D-Wave’s are well-adapted to these
problems, as they do not require the running of every combination of sequences
to obtain the lowest result. 
The D-Wave quantum computer has been used to solve a wide range of problems across various complexity classes such as training machine learning models \cite{date2020classical,arthur2021balanced,date2021qubo}, protein folding \cite{babej2018coarse}, graph partitioning \cite{ushijima2017graph} and portfolio optimization \cite{phillipson2021portfolio}.  
Combined with our work, optimal workflow paths may
be determined when a multitude of jobs are queued. On cloud resources like
IBMQ, this may help reduce wait times, while offering a practical
application for secondary quantum systems.

Researchers at ORNL and Alpine Quantum Technologies recently published a work
outlining the feasibility of HPC-Quantum Computing hybrid processes in
computing centers~\cite{9537178}. The outcomes reinforce that the hardware and
software for HPC and quantum computers are compatible with current technology,
and that the most significant limitation is in the quantum hardware, which
suffers from scaling and qubit coherence issues. 

Lastly, in the industry, a few companies are already committed to making changes in the
HPC world to implement a hybrid HPC and quantum computing environment. One of
the leaders in this field is Finland's IQM, one of Europe's leading quantum
computing firms. In their two part series ``The Way Forward: Bringing HPC and
Quantum Computing Together", they offer a three step process to combine current
HPC hardware and quantum computing resources~\cite{berebichez_2022}. The first
step involves identifying ways that quantum algorithms can optimize HPC
workflows, such as in quantum simulations, weather tracking and prediction
software, or optimization problems. The second and third
steps look at mid and long term goals, for example, designing the systems
needed based off of the research goals and finally implementing a workflow
between the supercomputers and quantum computer. They propose that the best way
to ensure the smooth flow of the process is to have HPC devices and quantum
computers in the same location, helping with troubleshooting and creating
further interest in the quantum capabilities the center has access to.

\section{Conclusions and Future Work}
\label{sec:concl}

We demonstrate the practicality of combining HPC and remote quantum resources for certain applications that require both resources. 
We use the Parsl workflow manager as we found it most conveniently adapted to the python platform. However, a variety of workflow platforms are available and we believe the same results may be achieved with any modern scientific workflow management system. In other words, our work is agnostic to specific solutions used and a validation for the workflows paradigm in general. Quantum devices are evolving and will likely act as auxiliary processors alongside the CPU and GPU. New low-level APIs to use such devices might be developed in the future. Currently, though, workflow systems offer a familiar and promising approach to combining the two.

In the future, we would like to test the same or similar algorithms on Quantinuum and Rigetti devices. The algorithms for both Shor's and the TSP required more credits than we were allotted for the project. Thus, we switched to IBMQ's resources, of which we had a significantly larger quota. If we had access to Quantinuum or Rigetti fully, it could offer insight into running workflows with changes in language. While IBMQ's quantum computers all take code in Qiskit, Quantinuum uses QASM and Rigetti uses pyQuil. Offering a workflow that could convert a quantum circuit from any one of these languages to the others would allow users much greater access to quantum machines, from Quantinuum's H1-1 to Rigetti's Aspen QPU. As research in Quantum Error Correction has peaked in recent years, we may assume further optimization in new quantum processors, allowing for larger circuits to run with less error, creating results more on par with the simulators used in this project.

\medskip
{\small
\noindent \textbf{\emph{Acknowledgments.}}
We acknowledge the use of IBM Quantum services for this work. The views expressed are those of the authors, and do not reflect the official policy or position of IBM or the IBM Quantum team. This research used resources of the Oak Ridge Leadership Computing Facility, which is a DOE Office of Science User Facility supported under Contract DE-AC05-00OR22725. 
}

\bibliographystyle{ieeetr}
\bibliography{references}

\begin{thebibliography}{10}

\bibitem{moores-law}
G.~E. Moore, ``{Cramming more components onto integrated circuits, Reprinted from Electronics, volume 38, number 8, April 19, 1965, pp.114 ff.},'' {\em IEEE Solid-State Circuits Society Newsletter}, vol.~11, no.~3, pp.~33--35, 2006.

\bibitem{end-of-moores-law}
T.~N. Theis and H.-S.~P. Wong, ``{The End of Moore's Law: A New Beginning for Information Technology},'' {\em Computing in Science \& Engineering}, vol.~19, no.~2, pp.~41--50, 2017.

\bibitem{ferreiradasilva2021works}
R.~Ferreira~da Silva, H.~Casanova, K.~Chard, I.~Altintas, R.~M. Badia, {\em et~al.}, ``{A Community Roadmap for Scientific Workflows Research and Development},'' in {\em 2021 IEEE Workshop on Workflows in Support of Large-Scale Science (WORKS)}, pp.~81--90, 2021.

\bibitem{ibmroadmap}
``{The IBM Quantum Development Roadmap}.'' \url{https://www.ibm.com/quantum/roadmap}.
\newblock Accessed: 2023-02-01.

\bibitem{Bieberich_HPC-QC-Workflows_2022}
S.~Bieberich, ``{HPC-QC-Workflows}.'' \url{https://github.com/Sam-Bieberich/HPC-QC-Workflows}, 8 2022.

\bibitem{crusher}
``Crusher.'' \url{https://docs.olcf.ornl.gov/systems/crusher_quick_start_guide.html}, 2022.

\bibitem{junger1995traveling}
M.~J{\"u}nger, G.~Reinelt, and G.~Rinaldi, ``{The Traveling Salesman Problem},'' {\em Handbooks in operations research and management science}, vol.~7, pp.~225--330, 1995.

\bibitem{hagberg2008exploring}
A.~Hagberg, P.~Swart, and D.~S~Chult, ``{Exploring network structure, dynamics, and function using NetworkX},'' tech. rep., Los Alamos National Lab.(LANL), Los Alamos, NM (United States), 2008.

\bibitem{johnston_harrigan_gimeno_segovia_2019}
E.~R. Johnston, N.~Harrigan, and M.~Gimeno-Segovia, {\em Quantum Phase Estimation}, p.~155–169.
\newblock O'Reilly, 2019.

\bibitem{bergamaschi_2020}
T.~R.-W. Bergamaschi, ``{Quantum Approximate Optimization Algorithms on the Traveling Salesman Problem},'' {\em Medium}, Feb 2020.

\bibitem{radzihovsky_murphy_swofford_2019}
M.~Radzihovsky, J.~Murphy, and M.~Swofford, ``{A QAOA solution to the traveling salesman problem using pyQuil},'' {\em stanford.edu}, May 2019.

\bibitem{parsl}
Y.~Babuji {\em et~al.}, ``{Parsl: Pervasive Parallel Programming in Python},'' in {\em Proceedings of the 28th International Symposium on High-Performance Parallel and Distributed Computing}, HPDC '19, (New York, NY, USA), p.~25–36, Association for Computing Machinery, 2019.

\bibitem{adedoyin2018quantum}
A.~Adedoyin {\em et~al.}, ``{Quantum Algorithm Implementations for Beginners},'' {\em arXiv preprint arXiv:1804.03719}, 2018.

\bibitem{Qiskit-Textbook}
A.~Abbas {\em et~al.}, ``{Learn Quantum Computation Using Qiskit},'' 2020.

\bibitem{Karlsson20184qubitGA}
V.~B. Karlsson and P.~Str{\"o}mberg, ``{4-qubit Grover's algorithm implemented for the ibmqx5 architecture},'' 2018.

\bibitem{Qiskit}
M.~S. ANIS {\em et~al.}, ``{Qiskit: An Open-source Framework for Quantum Computing},'' 2021.

\bibitem{amico2019experimental}
M.~Amico, Z.~H. Saleem, and M.~Kumph, ``{Experimental study of Shor's factoring algorithm using the IBM Q Experience},'' {\em Physical Review A}, vol.~100, no.~1, 2019.

\bibitem{10.1007/978-3-030-50433-5_40}
D.~Tomasiewicz, M.~Pawlik, M.~Malawski, and K.~Rycerz, ``{Foundations for Workflow Application Scheduling on D-Wave System},'' in {\em Computational Science -- ICCS 2020}, (Cham), pp.~516--530, Springer International Publishing, 2020.

\bibitem{10.1007/978-3-030-77980-1_10}
D.~Vietz, J.~Barzen, F.~Leymann, and K.~Wild, ``{On Decision Support for Quantum Application Developers: Categorization, Comparison, and Analysis of Existing Technologies},'' in {\em Computational Science -- ICCS 2021}, (Cham), pp.~127--141, Springer International Publishing, 2021.

\bibitem{weber_2022}
B.~Weber, ``{QuantMe}.'' https://github.com/UST-QuAntiL/QuantME-TransformationFramework, Aug 2022.

\bibitem{distributed}
W.~Bo{\.{z}}ejko, J.~Pempera, M.~Uchro{\'{n}}ski, and M.~Wodecki, ``{Distributed Quantum Annealing on D-Wave for the Single Machine Total Weighted Tardiness Scheduling Problem},'' in {\em Computational Science -- ICCS 2022}, (Cham), pp.~171--178, Springer International Publishing, 2022.

\bibitem{date2020classical}
P.~Date, C.~Schuman, R.~Patton, and T.~Potok, ``A classical-quantum hybrid approach for unsupervised probabilistic machine learning,'' in {\em Advances in Information and Communication: Proceedings of the 2019 Future of Information and Communication Conference (FICC), Volume 2}, pp.~98--117, Springer, 2020.

\bibitem{arthur2021balanced}
D.~Arthur and P.~Date, ``Balanced k-means clustering on an adiabatic quantum computer,'' {\em Quantum Information Processing}, vol.~20, pp.~1--30, 2021.

\bibitem{date2021qubo}
P.~Date, D.~Arthur, and L.~Pusey-Nazzaro, ``Qubo formulations for training machine learning models,'' {\em Scientific reports}, vol.~11, no.~1, p.~10029, 2021.

\bibitem{babej2018coarse}
T.~Babej, M.~Fingerhuth, {\em et~al.}, ``Coarse-grained lattice protein folding on a quantum annealer,'' {\em arXiv preprint arXiv:1811.00713}, 2018.

\bibitem{ushijima2017graph}
H.~Ushijima-Mwesigwa, C.~F. Negre, and S.~M. Mniszewski, ``Graph partitioning using quantum annealing on the d-wave system,'' in {\em Proceedings of the Second International Workshop on Post Moores Era Supercomputing}, pp.~22--29, 2017.

\bibitem{phillipson2021portfolio}
F.~Phillipson and H.~S. Bhatia, ``{Portfolio Optimisation using the D-Wave Quantum Annealer},'' in {\em Computational Science--ICCS 2021: 21st International Conference, Krakow, Poland, June 16--18, 2021, Proceedings, Part VI}, pp.~45--59, Springer, 2021.

\bibitem{9537178}
T.~S. Humble, A.~McCaskey, D.~I. Lyakh, M.~Gowrishankar, A.~Frisch, and T.~Monz, ``{Quantum Computers for High-Performance Computing},'' {\em IEEE Micro}, vol.~41, pp.~15--23, sep 2021.

\bibitem{berebichez_2022}
D.~Berebichez, ``{The way forward: Bringing HPC and Quantum Computing together (part 1 \& 2)}.'' \url{https://www.meetiqm.com/articles/blog/the-way-forward}, Apr 2022.

\end{thebibliography}

\vspace{12pt}
\color{red}
\end{document}